\documentclass[a4paper]{jpconf}
\usepackage{amsthm,amsfonts,graphicx,verbatim}
\usepackage{graphicx}
\usepackage{amsmath}
\usepackage{amssymb}
\usepackage{amsthm}
\usepackage{amsfonts}
\usepackage{listings}

\newcommand{\be}{\begin{equation}}
\newcommand{\ee}{\end{equation}}
\newcommand{\bea}{\begin{eqnarray}}
\newcommand{\eea}{\end{eqnarray}}

\begin{document}
\title{Numerical studies of the fractional quantum Hall effect in systems with tunable interactions}

\author{Z. Papi\'c$^1$, D. A. Abanin$^2$, Y. Barlas$^3$, and R. N. Bhatt$^{1,2}$}
\address{$^1$ Department of Electrical Engineering, Princeton University, Princeton, NJ 08544, USA}
\address{$^2$ Princeton
 Center for Theoretical Science, Princeton University, Princeton, NJ
 08544, USA} 
\address{$^3$ National High Magnetic Field Laboratory and Department of Physics, Florida State University, FL 32306, USA}


\begin{abstract}
The discovery of the fractional quantum Hall effect in GaAs-based semiconductor devices has lead to new advances in condensed matter physics, in particular the possibility for exotic, topological phases of matter that possess fractional, and even non-Abelian, statistics of quasiparticles. One of the main limitations of the experimental systems based on GaAs has been the lack of tunability of the effective interactions between two-dimensional electrons, which made it difficult to stabilize some of the more fragile states, or induce phase transitions in a controlled manner. Here we review the recent studies that have explored the effects of tunability of the interactions offered by alternative two-dimensional systems, characterized by non-trivial Berry phases and including graphene, bilayer graphene and topological insulators. The tunability in these systems is achieved via external fields that change the mass gap, or by screening via dielectric plate in the vicinity of the device. Our study points to a number of different ways to manipulate the effective interactions, and engineer phase transitions between quantum Hall liquids and compressible states in a controlled manner.    
\end{abstract}

\section{Introduction}

Strongly correlated phases of electrons confined to move in the plane, subject to a perpendicular magnetic field,
have attracted significant attention since the discovery of fractionally quantized Hall conductivity~\cite{tsg}. 
The profound role of topology in this extreme quantum limit leads to the presence of quasiparticles that carry
a fraction of electron charge~\cite{laughlin}, and fractional (Abelian, or possibly non-Abelian) statistics~\cite{laughlin, halperin_statistics, asw_statistics, mr, nw_statistics, bgn_statistics}.
The prospect of excitations possessing non-Abelian statistics has motivated different schemes for topological quantum computation~\cite{tqc}
based on these systems.

These remarkable phenomena occur in the fractional quantum Hall (FQH) regime, when the number of electrons, $N_e$, 
is comparable to the number of magnetic flux quanta $N_\Phi$ through the two-dimensional electron system (2DES).
Correlated FQH liquid states appear at certain partial filling $\nu = N_{e}/N_{\Phi}$ of the ``active" Landau level (LL).
In traditional semiconductor heterostructures, the physics of a partially-filled $n=1$ LL differs significantly
from that of $n=0$ LL, due to the node in the single-particle wavefunction~\cite{prange}. As a consequence, the
hierarchy/composite fermion states~\cite{Haldane86, jainbook}, ubiquitous in the lowest Landau level (LLL), are significantly weakened in 
$n=1$ LL, and some of the more exotic states, such as the Read-Rezayi (RR) parafermion states~\cite{rr_parafermion}, 
are likely to be favored. A number of studies have focused on the simplest, $k=2$ non-Abelian member of the RR sequence --
the Moore-Read (MR) ``Pfaffian" state~\cite{mr}, believed to describe the FQH plateau at $\nu=5/2$~\cite{eover4}.
Quasiparticles of the MR and higher RR states obey the non-Abelian statistics~\cite{mr} which is of interest for
topological quantum computation~\cite{tqc}. 

One of the main disadvantages of the GaAs-based devices is that their 2DES is buried inside a larger, three-dimensional
structure. This unfortunate fact fixes the effective interactions at values that are often not optimal for some of the most
interesting FQH states, including the RR series. For example, the MR state is found to lie very close to the boundary with
a compressible phase~\cite{Morf98, Rezayi00}. Another problem stems from strong dielectric screening and finite well-width~\cite{zds} in GaAs, 
which weaken the electron-electron interactions and make FQH states fragile. This has been a major obstacle in the studies of the non-Abelian states,
which could only be observed in ultra-high-mobility samples~\cite{eover4}. 
Thus, it is desirable to find an alternative high-mobility 2DES with strong effective Coulomb interactions that are adjustable in a broad range.

Recently, a new class of such high-mobility 2DES which host chiral excitations with non-trivial Berry phases, has been discovered. 
These ``chiral" materials include graphene and bilayer graphene~\cite{CastroNeto09}, and, more recently, topological insulators~\cite{KaneHasan},
as well as certain quantum wells~\cite{HgTe}. The chiral nature of the quasiparticles gives rise to new electronic properties, 
including an unusual LL sequence, the anomalous Hall effect, and suppression of weak localization~\cite{CastroNeto09}.
When these chiral materials are subject to a perpendicular magnetic field, the kinetic energy quenches into discrete Landau levels,
similar to the usual semiconductors with non-chiral carriers. However, due to the chiral band structure~\cite{massivedirac} and the fact that the surface
of these materials is exposed~\cite{tunable}, they offer new possibilities to tune the effective interactions and explore strongly correlated phases. 

In this paper we review two practical ways~\cite{tunable, massivedirac} of tuning the effective interactions in chiral 2DES, and the effect this has on the  
FQH states. One of the attractive features of the chiral 2DESs is that they allow 
for a more robust realization of certain FQH states in \emph{several} LLs, as opposed to a single LL in GaAs. 
Additional insights can be obtained by driving transitions between the incompressible (FQH) states and the compressible, stripe and bubble, phases. 
Such transitions can be implemented in chiral and \emph{massive} 2DESs by varying 
the external field. Overall, the tunability of the chiral 2DES allows one to explore a larger region of the effective interactions
than has been achieved in GaAs. 

\section{Model} \label{sec_model} 

We begin this Section by providing a brief overview of the one-body problem in a magnetic field, as well as the summary of the standard recipe of exact diagonalization applied to quantum Hall systems.  

Consider a single electron moving in a plane, subject to a perpendicular magnetic field $B\hat{z}=\nabla\times \mathbf{A}(\mathbf{r})$. The corresponding Hamiltonian can be written in a ``covariant" notation as
\begin{eqnarray}
K =\frac{1}{2m} g^{ab} \pi_{a} \pi_{b},
\end{eqnarray}
where $\pi_a= p_a - \frac{e}{c} A_a(\mathbf{r})$ ($a=x,y$) represents the dynamical momentum, and $\hat{g}$ is the band mass tensor. In the usual isotropic case ($g=I$), we can obtain the single-particle energies (Landau levels) by choosing, for example, the symmetric gauge $A_x=By/2, A_y=-Bx/2$. In this case, the dynamical momenta become 
$\pi_x=-i\hbar \frac{\partial}{\partial x}+\frac{\hbar}{2\ell_B^2}y$ and $\pi_y=-i\hbar \frac{\partial}{\partial y} - \frac{\hbar}{2\ell_B^2}x$, in terms of the magnetic length $\ell_B=\sqrt{\hbar/eB}$. Hamiltonian can be transformed into a diagonal form $K=\frac{\hbar\omega_c}{2} \left(a^\dagger a + \frac{1}{2} \right)$ with the help of ``ladder" operators 
\begin{eqnarray}\label{a_operators}
a \propto \pi_x + i\pi_y, \\
a^\dagger \propto \pi_x - i \pi_y.
\end{eqnarray}
However, for each value of $a^\dagger a$, there remains a residual degeneracy equal to the number of magnetic flux quanta, $N_\Phi$. This degeneracy is resolved by another pair of operators $b, b^\dagger$ that commute with $a, a^\dagger$ and depend on the \emph{guiding center} coordinates of the electron, $R^a = r^a - \frac{\epsilon^{ab}}{\hbar} \pi_b \ell_B^2$. Operators $b^\dagger$ create the (unnormalized) single particle eigenstates of the lowest, $n=0$ Landau level (LLL), 
\begin{equation}\label{singleparticleorbital}
\phi_l (z) \propto z^l e^{-z^* z/4\ell_B^2},
\end{equation}
with $z=x+iy$ being the complex coordinate of an electron in the plane (and $z^*$ denoting its complex-conjugate). The quantum number $l$ is an eigenvalue of the angular momentum, and the single particle states are localized along concentric rings around the origin. Single-particle states in higher LLs are constructed by the repeated action of $a^\dagger$, thus completing the solution of the one-body problem in a rotationally-invariant case.

\subsection{Many-body problem: Exact diagonalization}

An interacting problem of a finite number of electrons can be studied numerically, using exact diagonalization which has been remarkably successful in unravelling many of the essential physical properties of FQH systems, for systems as small as 10 electrons~\cite{prange,Haldane86}. A crucial approximation that makes exact diagonalization practical is to neglect the excitations between LLs and restrict the Hilbert space to a single, ``active" LL, which has a degeneracy of $N_\Phi$. This approximation is physically justified in high magnetic fields where such excitations indeed become very costly. One can then construct a many-body Hilbert space, consisting of Slater determinants $|l_1,l_2,\ldots\rangle$ built from the states (\ref{singleparticleorbital}), using a suitable choice of boundary condition.  
In the study of the FQH effect, two kinds of surfaces are available that preserve the translational invariance of an
infinite 2DES: sphere~\cite{Haldane86} and torus~\cite{pbc}. The two choices of boundary conditions
illustrate the specific features of a many-body FQH state under investigation: on a sphere, the FQH state couples to the curvature of the manifold, which is characterized by the topological number called shift~\cite{shift}. The shift produces a small offset between $N_\Phi=N/\nu$ and the magnetic monopole, whose strength is denoted by $2S$, placed in the center of a sphere. As a consequence, different FQH states ``live" in different Hilbert spaces and in principle can be directly compared only after extrapolation to the thermodynamic limit. On the other hand, the flat surface of a torus leads to a unique definition of $N_\Phi$ for given $N$ and $\nu$, such that different candidate states describing the same filling $\nu$ are all realized in the same Hilbert space. The caveats of this geometry are the additional geometric parameters, the angle and aspect ratio of the torus; the Hamiltonian of a finite-size system depends on these parameters and their specific values favor one FQH phase over the others. Thus, an analysis is slightly more involved but the gain is that the ground-state degeneracy can be used to identify topologically-ordered states. In contrast, on the sphere FQH states always appear as non-degenerate, zero angular momentum ground states. 

Once the Hilbert space is defined and properly adapted to the symmetry group/boundary condition, the remaining task is to represent the Hamiltonian and diagonalize the corresponding matrix using a variant of Lanczos algorithm~\cite{lanczos} (or Jacobi/Householder routine~\cite{numrec} for very small systems). The advantage of symmetry, in particular rotational invariance, is that Wigner-Eckart theorem applies, and leads to the definition of the so-called Haldane pseudopotentials~\cite{prange}. Any two-body interaction, such as Coulomb potential projected to a single LL, can be written in the usual second-quantized form
$H=\sum_{\{m_i\}} V_{m_1m_2m_3m_4} c_{m_1}^\dagger c_{m_2}^\dagger c_{m_4} c_{m_3}$. Furthermore, assuming rotational invariance, 
the matrix element factorizes into
\begin{equation}
V_{m_1m_2m_3m_4} = \sum_{L=0}^{2S} V_L \sum_{M=-L}^{L} C_{m_1m_2M}^{SSL} C_{m_3m_4M}^{SSL}, 
\end{equation}
i.e. into a product of a geometrical factor involving the Clebsch-Gordan coefficients $C^{J_1J_2J}_{m_1m_2m}$, and the Haldane pseudopotential $V_L$. The finite set of Haldane pseudopotentials contains all the information about the interaction projected to a single LL and thus determines all the physical properties of a FQH system.
Haldane pseudopotentials can be conveniently evaluated from the Fourier transform of the interaction, $V(q)$. 

\subsection{Modification of the interaction due to the spinor nature of the wavefunction in massive Dirac materials}

Once the Hilbert space is defined and symmetry-adapted, the Coulomb Hamiltonian can be represented in the matrix form. In the LLL, the interaction is simply given by $V(q)=1/q$, which is the Fourier transform of the Coulomb potential in two dimensions. However, if we consider the case when the chemical potential is such that LLs from 1 through $n-1$ are completely filled and ``inert", while the $n$th LL is partially filled and ``active", the effective interaction becomes slightly modified and assumes the form $1/q \times |F_n(q)|^2$, where the \emph{form factor}~\cite{prange} $F_n(q) = L_n(q^2\ell_B^2/2)$ can be evaluated using the algebra of operators $a,a^\dagger$, Eq. (\ref{a_operators}) -- see, e.g., Ref.~\cite{jainbook} ($L_n$ is the $n$th Laguerre polynomial). Similarly, if the 2DES has a finite extent in the $z$-direction, the interaction becomes modified via an appropriate form-factor that results from integrating out the $z$-component of the wavefunction.

The modification of the effective interaction may also result from the spinor nature of the wavefunction~\cite{Nomura06}. Such wavefunctions arise naturally in a family of 2D materials that have non-trivial Berry phases. One such material is monolayer graphene, a high-mobility atomically thick 2DES~\cite{CastroNeto09}, where recently several FQH states of the type $\nu=m/3$ have been discovered~\cite{graphene_fqhe}.  A closely
related material, bilayer graphene~\cite{CastroNeto09} has similarly high mobility, and exhibits interaction-induced quantum Hall states at integer filling factors at low magnetic fields~\cite{feldman-09np889}. Graphene and its bilayer are characterized by Berry phase $\pi$ (graphene-like) and $2\pi$ (bilayer-graphene-like with an energy gap), respectively. More explicitly, we refer to the Dirac materials as those described by a family of $2\times 2$ Hamiltonians with Berry phase $\pi$ and $2\pi$~\cite{massivedirac}. The case of $\pi$-carriers is realized in graphene, topological insulators, special quantum wells~\cite{HgTe}; the case of $2\pi$-carriers occurs in bilayer graphene, where the energy gap can be controlled in a wide range by a perpendicular electric field~\cite{CastroNeto09}. Details on how the effective $2\times 2$ Hamiltonian can be derived from a more complete tight-binding model can be found in the review~\cite{barlas_review}.

For $\pi$ carriers, the single-particle wavefunctions are given by spinors~\cite{massivedirac}
$\psi_n=(\cos\theta_n \phi_{|n|-1},\sin\theta_n \phi_{|n|})$, where $\phi_n$ is the wave function of the $n$th non-relativistic LL ($n=\pm 1, \pm 2, \ldots$), and parameter $\theta$ depends on the ratio $\Delta/(\hbar v_0/\ell_B)$, where $\Delta$ is the mass gap, and $v_0$ the Fermi velocity. We use the notation $(\sigma\pi,n)$ to denote the $n$th LL for $\sigma\pi$ carriers ($\sigma=1,2$).  As a consequence of the spinor wavefunction, the effective form factor~\cite{prange} $F_{n}^{\pi} (q)$ that describes the interaction projected to a $(\pi,n)$ LL is given by
\begin{equation}\label{eq:formfactors}
F_{n}^{\pi} (q)=\cos^2\theta L_{|n|-1} \left(\frac{q^2 \ell_B^2}{2}\right)+\sin^2\theta L_{|n|}\left(\frac{q^2 \ell_B^2}{2}\right),
\end{equation}
where for simplicity we omitted the index of $\theta$. The form-factor is a mixture of the $(|n|-1)$th and $|n|$th LL form-factors in a non-relativistic 2DES with parabolic dispersion. At  $\theta=\pi/4$, the above equation reduces to the form-factor of graphene~\cite{Nomura06}, however, by varying $\Delta/(\hbar v_0/\ell_B)$ in the experimentally accessible range~\cite{massivedirac}, one can realize any value of $\theta\in (0;\pi/2)$. 

Similarly, for carriers with Berry phase $2\pi$, the single-particle wavefunctions are $\psi_n=(\cos\theta_n \phi_{|n|-1}, \sin\theta_n \phi_{|n|+1} )$, and the form-factor is a mixture of standard $(|n|-1)$th and $(|n|+1)$th form-factors, 
\begin{equation}\label{eq:formfactors2}
F_{n}^{2\pi} (q)=\cos^2\theta L_{|n|-1} \left(\frac{q^2\ell_B^2}{2}\right)+\sin^2\theta L_{|n|+1}\left(\frac{q^2 \ell_B^2}{2}\right). 
\end{equation}

\begin{center}
\begin{figure}[t]
\centerline{\includegraphics[scale=0.85]{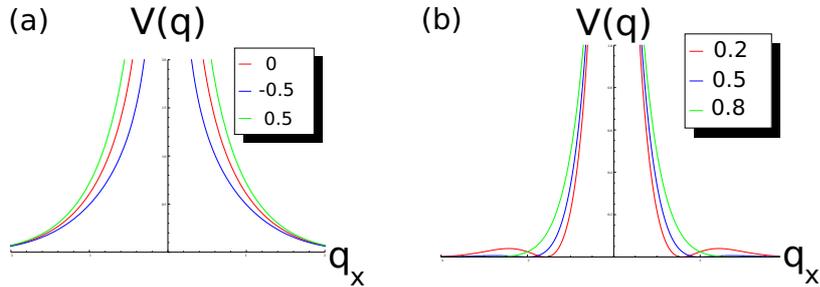}}
\caption[]{(Color online) Two ways of modifying the effective interaction $V(q)$ in the FQH regime: (a) via dielectric screening, for several values of $\alpha$ given in the inset (using Eq.(\ref{eq:interaction_screened})) for $d=\ell_B$, (b) via chiral band structure, for several values of $\cos^2\theta$ given in the inset (in $(\pi,1)$ LL). Because the interaction $V(q)$ is isotropic, it is plotted as a function of $q_x$ only. The comparison of the corresponding Haldane pseudopotentials for cases (a) and (b) can be found in Ref.~\cite{papic_njp}.}
\label{fig:pp}
\end{figure}
\end{center}

The tunable form of the effective interactions Eqs.(\ref{eq:formfactors},\ref{eq:formfactors2}) provides a way to engineer transitions between strongly correlated phases {\it in situ} by changing the field, as illustrated in Fig.~\ref{fig:pp}(b).

\subsection{Modification of the effective interaction due to dielectric screening}

Up to this point, we have discussed the change in the effective interaction $V(q)|F(q)|^2$ that resulted from a modification of the single-particle wavefunctions, and therefore $F(q)$. The method, proposed in Ref.~\cite{tunable}, allows one to directly change $V(q)$. Chiral materials, such as graphene, are often exposed to the environment, which allows for dielectric material to be deposited on top of them. We consider a setup where graphene sample is situated in a dielectric medium with permittivity $\epsilon_1$, and a semi-infinite dielectric plate with permittivity $\epsilon_2\neq \epsilon_1$ is placed at a distance
$d/2$ away from the graphene sheet. The effective interactions between electrons in graphene change due to the surface charges induced at the boundary
between dielectrics~\cite{tunable}:
\begin{equation}\label{eq:interaction_screened}
V(r)=\frac{e^2}{\epsilon_1 r}+\alpha \frac{e^2}{\epsilon_1\sqrt{r^2+d^2}}, \,\, {\rm where} \, \,
\alpha=\frac{\epsilon_1-\epsilon_2}{\epsilon_1+\epsilon_2}.
\end{equation}
The ratio $d/\ell_B$ controls the effective interactions within a partially filled LL. However, the overall energy scale is also modified and this has an impact on the magnitude of the excitation gap. The gap should be multiplied by a factor $\epsilon_{\rm GaAs}/\epsilon_1$ if comparison is to be made with GaAs 2DES. Again, an important advantage of this setup is that the interactions can be tuned {\it in situ} by varying the magnetic field $B$, which modifies the ratio $d/\ell_B$. The consequences for the many-body system are illustrated in Fig.~\ref{fig:pp}(a).

\section{Results}\label{sec_results}

We now present some exact-diagonalization results for the interaction models introduced in Sec.\ref{sec_model}. The focus is on the physical features resulting from the variation of the effective interaction; as stated in the introduction, we follow the traditional assumption of neglecting the excitations between different LLs, which is expected to be quite reasonable in GaAs, but perhaps less so in chiral materials. Much of previous theoretical work on chiral materials has been restricted to graphene, exploring the consequences of the four-fold LL degeneracy (valley and spin) that leads to new SU(2) and SU(4)-symmetric fractional states~\cite{multi-component}. Instead, we focus on the high-field limit, neglecting the multicomponent degrees of freedom, and examine the effects originating from the interplay of the Coulomb interaction and band structure. We consider a family of band structures introduced in Ref.~\cite{massivedirac}, which describe a number of high-mobility materials, including graphene with \emph{massive} carriers (mass is generated either spontaneously, or as a result of sublattice symmetry breaking~\cite{CastroNeto09}), topological insulators~\cite{KaneHasan}, bilayer-graphene~\cite{CastroNeto09}, trilayer graphene~\cite{trilayer}, and similar materials. Pristine graphene, which hosts massless Dirac-like fermions, is contained in this model as a particular case. Note that we explore fairly large variations of the effective interactions among fully polarized electrons; more subtle effects due to LL mixing~\cite{llmixing} and multicomponent degrees of freedom are left for future study.

We show the results for both spherical and torus geometry. A number of useful insights can be inferred from the study of the energy spectrum; in addition, we use overlap calculations to compare an exact, many-body ground state $\Psi_{\rm exact}$, with a numerical representation of a trial wavefunction, $\Psi_{\rm trial}$. The overlap is defined as a scalar product between two normalized vectors, $\mathcal{O}=|\langle \Psi_{\rm trial} | \Psi_{\rm exact} \rangle |$. If $\mathcal{O}$ is consistently close to unity for a number of system sizes considered, we consider the trial wavefunction to be a faithful description of a FQH phase. From the knowledge of a ground-state wavefunction, we also evaluate the (projected) static structure factor~\cite{sq}. Sharp peaks in the structure factor indicate the onset of compressible phases~\cite{stripe_bubble, stripe_bubble_numerics}.

\begin{figure}[t]
\centerline{\includegraphics[scale=0.5]{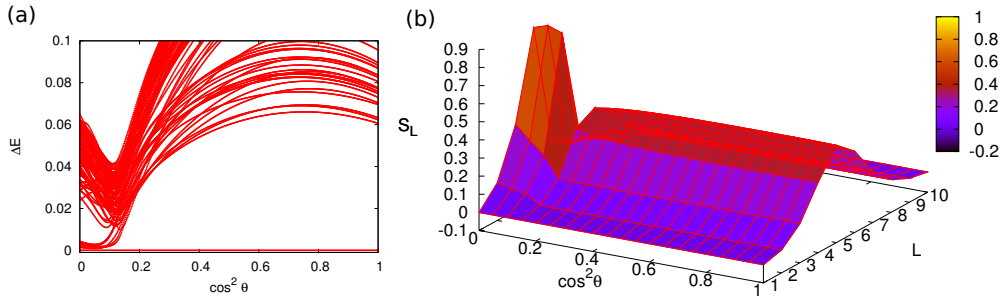}}
\caption[]{(Color online) Phase transition between the Laughlin and bubble phase in $(2\pi,1)$ LL. (a) Low-lying energy spectrum on the torus as a function of $\theta$. (b) Projected structure factor $S_L$ as a function of angular momentum $L$ on the sphere and parameter $\theta$. Large ground-state degeneracy for small $\cos^2\theta$, as well as the sharp peak in $S_L$, correspond to the bubble phase.}
\label{fig_bill0}
\end{figure}
The simplest example of the phase transition can be realized between the Laughlin state and a bubble phase in $(2\pi,1)$ LL of the chiral materials. In Fig. \ref{fig_bill0}(a) we show the low-lying part of the energy spectrum obtained by exact diagonalization in the torus geometry, as a function of parameter $\theta$ introduced in Sec.~\ref{sec_model}. For values $\cos^2\theta \geq 0.2$, the ground state of the system is non-degenerate and has a large overlap with the Laughlin wavefunction. When $\cos^2\theta$ is less than 0.2, the effective interaction resembles that of $n=2$ non-relativistic LL, which favors the bubble phase at filling $\nu=1/3$. This is manifested by a large ground-state degeneracy developing in the spectrum~\cite{stripe_bubble_numerics}. Similarly, the evolution of the projected structure factor $S_L$ on the sphere as a function of $\theta$ develops a discontinuity around $\cos^2\theta\approx 0.2$ and a sharp peak below that value, signaling the compressibility of the ground state.  

\begin{figure}[t]
\centerline{\includegraphics[scale=0.4]{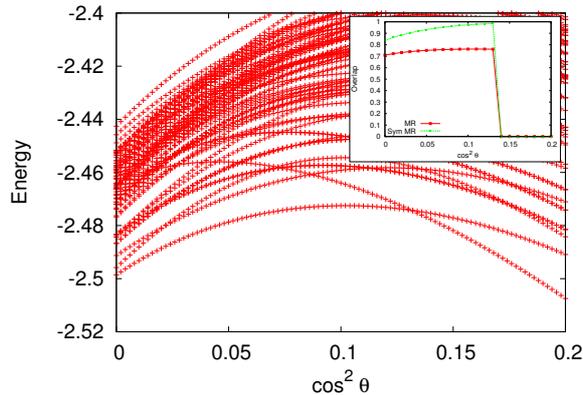}}
\caption[]{(Color online) Enhanced Pfaffian state and the transition to the CFL. Energy spectrum of $N=12$ electrons on the torus in $(\pi,1)$ LL is plotted as a function of $\theta$. Inset shows the ground-state overlap with the Moore-Read wavefunction and its particle-hole symmetrized version.}
\label{fig_pf}
\end{figure}

Previously~\cite{massivedirac}, it was emphasized that the tunability of the interactions in chiral materials can lead to the more stable non-Abelian states. One such example of considerable experimental interest is the Moore-Read Pfaffian state~\cite{mr}. In Fig.~\ref{fig_pf} we show the energy spectrum for $N=12$ electrons on a torus in $(\pi,1)$ LL. At $\cos^2\theta=0$, the interaction reduces to $n=1$ non-relativistic Coulomb interaction, which leads to a fragile Pfaffian state. As shown in the inset of Fig.~\ref{fig_pf}, the overlap of the exact ground state with the Moore-Read wavefunction slightly increases as $\cos^2\theta$ is increased from zero. However, due to the particle-hole symmetry which is present on the torus at filling $\nu=1/2$, a better indicator is to compare the exact ground state with a \emph{particle-hole symmetrized} version of the Moore-Read wavefunction~\cite{Rezayi00}. Indeed, in this case the overlap rises to nearly unity right before the transition at $\cos^2\theta\approx 0.14$. At this point, the level crossing occurs, and a composite Fermi liquid (CFL) state becomes the ground state of the system. We emphasize that although the level crossing would suggest a first-order transition, we cannot reliably rule out a possibility for the second-order transition in the thermodynamic limit. Note that, in addition to the considerable increase in the overlap with the Moore-Read state, also the excitation gap increases towards the transition to the CFL state, adding to the overall stability of the Pfaffian. As shown in Ref.~\cite{massivedirac}, Pfaffian correlations can also be realized in $(\pi,2)$ LL, and other states such as $k=3$ RR state can be realized in several LLs~\cite{papic_njp}.

\begin{figure}[t]
\centerline{\includegraphics[scale=0.9]{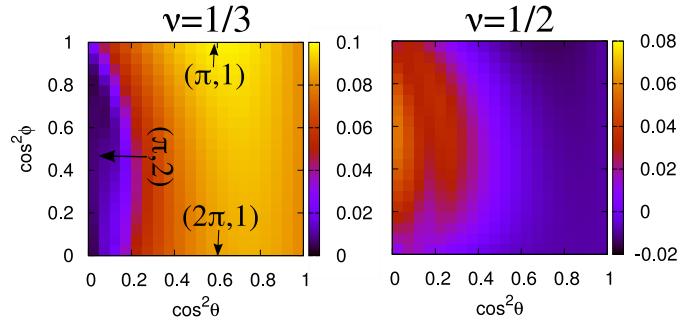}}
\caption[]{(Color online) Charge gap (in units of $e^2/\epsilon \ell_B$) at $\nu=1/3$ and $\nu=1/2$ as a function of the band-structure parameters $\theta$ and $\phi$. The calculation is performed in the spherical geometry, for $N=10$ particles at $\nu=1/3$ and $N=14$ particles at $\nu=1/2$, with a fixed shift of $-3$.}
\label{fig_chargegap}
\end{figure}

Phase transitions can also be studied without reference to a particular wavefunction by evaluating the excitation gap of the system, Fig.~\ref{fig_chargegap}. To be more general, we consider a model that involves a superposition of $n=0$, $n=1$ and $n=2$ LL non-relativistic form factors~\cite{massivedirac}:
\begin{equation}\label{eq:formfactor_general}
F_n(q) = \cos^2\theta L_{|n|} \left(\frac{q^2 \ell_B^2}{2}\right)  +\sin^2\theta \cos^2\phi L_{|n|+1}\left(\frac{q^2 \ell_B^2}{2}\right)  +\sin^2\theta \sin^2\phi L_{|n|+2}\left(\frac{q^2 \ell_B^2}{2}\right).
\end{equation}
This model generalizes the case of $(\pi,1)$, $(\pi,2)$ and $(2\pi,1)$ LLs studied earlier, and naturally arises in systems such as trilayer graphene~\cite{trilayer}. In Fig. \ref{fig_chargegap} we plot the charge gap at $\nu=1/3$ and $\nu=1/2$ as a function of $\theta$ and $\phi$. Because the calculation is performed in the spherical geometry, we fix the shift to be $-3$ (corresponding to the Laughin state at $\nu=1/3$ and the Moore-Read state at $\nu=1/2$).
Along certain lines, indicated by arrows, the phase diagram reduces to one of the cases mentioned above. Note that proper finite-size scaling needs to be performed in order to get the correct values for the gap; however, it was previously found~\cite{tunable,massivedirac} that this rigorous analysis produces values that are roughly in agreement with the ones shown in Fig. \ref{fig_chargegap}.
At $\nu=1/3$ the Laughlin state is realized for a wide range of $\theta$ and $\phi$. The maximum of the gap occurs for the effective interaction that is a mixture of $n=0$ and $n=1$ non-relativistic form-factors. For $\cos^2\theta \leq 0.2$ the excitation gap is significantly reduced and a bubble phase is realized. The point of transition to the bubble phase is insensitive to the value of $\phi$, unless $\cos^2\phi \approx 1$. On the other hand, at $\nu=1/2$ the behavior of the gap suggests the existence of three phases. On the right side (large $\cos^2\theta$), the gaps are close to zero -- this is the region corresponding to the CFL phase. In this parameter regime, the ground state of the system also has a different shift from the one chosen in Fig.~\ref{fig_chargegap}. On the left side, one can discern two distinct branches -- the middle one corresponds to the Moore-Read state because the overlap with the Pfaffian wavefunction is high in this region~\cite{massivedirac}. Note that this region is fairly large and encompasses the effective interactions which are quite different from the $n=1$ non-relativistic Coulomb interaction, known to give rise to the Pfaffian state. On the far left there is a stripe phase; the excitation gap is seemingly large in this region, however this is only an artefact of the spherical geometry that cannot adequately accommodate the broken-symmetry state such as the stripe. We expect this gap to be significantly reduced as one approaches the thermodynamic limit. 
Note that the gaps of the incompressible states can be further increased using a dielectric setup proposed in Ref.~\cite{tunable}.

\section{Conclusion}

In this paper we have discussed two practical ways of tuning the effective interactions in fractional quantum Hall systems. We emphasized an important role of chiral 2DESs where the interactions can be varied in a broader range than in GaAs, and more robust non-Abelian states are likely to be found. Moreover, these non-Abelian states are expected to exist in the new regimes of the effective interactions that are significantly different from the ones in GaAs. Finally, their excitation gaps can be tuned, and quantum phase transitions can be engineered \emph{in situ} by varying the magnetic field.

\section{Acknowledgements} This work was supported by DOE grant DE-SC$0002140$. Y.B. was supported by the State of Florida.


\section*{References}

\begin{thebibliography}{9}


\bibitem{tsg}
D. C. Tsui, H. L. Stormer, and A. C. Gossard, Phys. Rev. Lett. {\bf 48}, 1559 (1982).

\bibitem{laughlin}
R. B. Laughlin, Phys. Rev. Lett. {\bf 50}, 1395 (1983).

\bibitem{halperin_statistics}
B. I. Halperin, Phys. Rev. Lett. {\bf 52}, 1583 (1984).

\bibitem{asw_statistics}
D. Arovas, J. R. Schrieffer, and Frank Wilczek, Phys. Rev. Lett. {\bf 53}, 722 (1984).

\bibitem{mr}
G. Moore and N. Read, Nucl. Phys. B {\bf 360}, 362 (1991).

\bibitem{nw_statistics}  
C. Nayak and F. Wilczek, Nucl. Phys. B {\bf 479}, 529 (1996).

\bibitem{bgn_statistics}
P. Bonderson, V. Gurarie, and C. Nayak, Phys. Rev. B {\bf 83}, 075303 (2011). 



\bibitem{tqc} C. Nayak \emph{et al.}, Rev. Mod. Phys. {\bf 80}, 1083 (2008).

\bibitem{prange}
\emph{The Quantum Hall Effect}, 2nd ed., edited by R. E. Prange and S. M. Girvin, Springer-Verlag, New York, 1990.

\bibitem{Haldane86}
F. D. M. Haldane, Phys. Rev. Lett. {\bf 51}, 605 (1983).

\bibitem{jainbook} J. K. Jain, \emph{Composite fermions}, (Cambridge University Press, 2007).



\bibitem{rr_parafermion}
N. Read and E. Rezayi, Phys. Rev. B {\bf 59}, 8084 (1999).

\bibitem{eover4}
R. Willett \emph{et al.}, Phys. Rev. Lett. {\bf 59}, 1776 (1987).


\bibitem{Morf98}
R. H. Morf, Phys. Rev. Lett. {\bf 80}, 1505 (1998).

\bibitem{Rezayi00}
E. H. Rezayi and F. D. M. Haldane, Phys. Rev. Lett. {\bf 84}, 4685 (2000).

\bibitem{zds} F. C. Zhang and S. Das Sarma, Phys. Rev. B {\bf 33}, 2903
(1986).

\bibitem{CastroNeto09}
A. H. Castro Neto {\it et al.}, Rev. Mod. Phys. {\bf 81}, 109 (2009).

\bibitem{KaneHasan}
M. Z. Hasan and C. L. Kane, Rev. Mod. Phys. {\bf 82}, 3045 (2010). 

\bibitem{HgTe}
B. A. Volkov and O. A. Pankratov, Pisma Zh. Eksp. Teor. Fiz. 42, 145; JETP Lett. {\bf 42}, 178  (1985); 
B. Buttner {\it et al.}, Nature Phys. {\bf 7}, 418 (2011).

\bibitem{massivedirac}
Z. Papi\'c, D. A. Abanin, Y. Barlas, and R. N. Bhatt, arXiv:1108.1339.

\bibitem{tunable}
Z. Papi\'c, R. Thomale, and D. A. Abanin, Phys. Rev. Lett. {\bf 107}, 176602 (2011). 

\bibitem{pbc}
D. Yoshioka, B. I. Halperin, and P. A. Lee, Phys. Rev. Lett. {\bf 50}, 1219 (1983); F. D. M. Haldane in Ref.~\cite{prange}.

\bibitem{shift}
X. G. Wen and A. Zee, Phys. Rev. Lett. {\bf 69}, 953 (1992).

\bibitem{lanczos}
J. Cullum and R. A. Willoughby, J. Comp. Phys. {\bf 44}, 329 (1981).

\bibitem{numrec}
W. H. Press, S. A. Teukolsky, W. T. Vetterling, and B. P. Flannery, \emph{Numerical recipes}, 2nd ed. (Cambridge University Press, Cambridge, 1992).

\bibitem{Nomura06}
K. Nomura and A. H. MacDonald, Phys. Rev. Lett. {\bf 96}, 256602 (2006); M. O. Goerbig, R. Moessner, and B. Dou\c{c}ot,
Phys. Rev. B {\bf 74}, 161407 (2006).

\bibitem{graphene_fqhe}
X. Du {\it et al.}, Nature {\bf 462}, 192 (2009);
K. Bolotin {\it et al.}, Nature {\bf 462}, 196 (2009);
C. R. Dean {\it et al.}, arXiv:1010.1179. 



\bibitem{feldman-09np889} B. E. Feldman, J. Martin, A. Yacoby, Nature Phys. {\bf 5}, 889 (2009).

\bibitem{barlas_review}
Y. Barlas, K. Yang, and A. H. MacDonald, arXiv:1110.1069


\bibitem{papic_njp}
D. A. Abanin, Z. Papi\'c, Y. Barlas, and R. N. Bhatt (submitted to New Journal of Physics)


\bibitem{multi-component}
V. M. Apalkov and T. Chakraborty, Phys. Rev. Lett. {\bf 97}, 126801 (2006); 
C. T\H{o}ke and J. Jain, Phys. Rev. B {\bf 75}, 245440 (2007); 
Z. Papi\'c, M. O. Goerbig, and N. Regnault, Phys. Rev. Lett. {\bf 105}, 176802 (2010).


\bibitem{trilayer}
W. Bao {\it et al.},  arXiv:1103.6088 (2011); T. Taychatanapat {\it et al.},  arXiv:1104.0438 (2011); A. Kumar {\it et al.},  arXiv:1104.1020 (2011); C. H. Lui {\it et al.},  arXiv:1105.4658 (2011). 


\bibitem{llmixing}
W. Bishara and C. Nayak, Phys. Rev. B {\bf 80}, 121302 (2009); A. Wojs, C. T\H{o}ke, and J. K. Jain, Phys. Rev. Lett. {\bf 105}, 096802 (2010); 
E. H. Rezayi and S. H. Simon, Phys. Rev. Lett. {\bf 106}, 116801 (2011).


\bibitem{stripe_bubble}
A. A. Koulakov, M. M. Fogler, and B. I. Shklovskii, Phys. Rev. Lett. {\bf 76}, 499 (1996); 
R. Moessner and J. T. Chalker, Phys. Rev. B {\bf 54}, 5006 (1996).


\bibitem{sq}
S. He, S. H. Simon and B. I. Halperin, Phys. Rev. B {\bf 50}, 1823 (1994).

\bibitem{stripe_bubble_numerics}
E. H. Rezayi, F. D. M. Haldane, and K. Yang, Phys. Rev. Lett. {\bf 83}, 1219 (1999);
F. D. M. Haldane, E. H. Rezayi, and K. Yang, Phys. Rev. Lett. {\bf 85}, 5396 (2000).





\end{thebibliography}
\end{document}